\documentclass[aps,prl,twocolumn,superscriptaddress,a4paper]{revtex4}

\usepackage{amsmath} 
\usepackage{graphicx}
\def\U#1{{%
\def\O{\mbox{O}}
\def\u{\mbox{u}}
\mathcode`\u=\mu
\mathcode`\O=\Omega
\mathrm{#1}}}
\def\ii{{\mathrm{i}}}
\def\ee{{\mathrm{e}}}

\def\ket#1{|\mbox{$#1$}\rangle}

\def\bracketi#1#2{\langle\mbox{$#1$}|\mbox{$#2$}\rangle}

\def\vct#1{{\mathchoice{\mbox{\boldmath$#1$}}{\mbox{\boldmath$#1$}}%
  {\mbox{\scriptsize\boldmath$#1$}}{\mbox{\scriptsize\boldmath$#1$}}}}

\begin{document}
\title{Direct observation of geometric phases using a three-pinhole interferometer}
\author{H. Kobayashi}
\affiliation{Department of Electronic Science and Engineering, Kyoto
University, Kyoto 615-8510, Japan}
\author{S. Tamate}
\affiliation{Department of Electronic Science and Engineering, Kyoto
University, Kyoto 615-8510, Japan}
\author{T. Nakanishi}
\affiliation{Department of Electronic Science and Engineering, Kyoto
University, Kyoto 615-8510, Japan}
\affiliation{CREST, Japan Science and Technology Agency, Saitama 332-0012,
Japan}
\author{K. Sugiyama}
\affiliation{Department of Electronic Science and Engineering, Kyoto
University, Kyoto 615-8510, Japan}
\affiliation{CREST, Japan Science and Technology Agency, Saitama 332-0012,
Japan}
\author{M. Kitano}
\affiliation{Department of Electronic Science and Engineering, Kyoto
University, Kyoto 615-8510, Japan}
\affiliation{CREST, Japan Science and Technology Agency, Saitama 332-0012,
Japan}
\date{\today}

\begin{abstract}
We present a method to measure the geometric phase defined for three
 internal states of a photon (polarizations) 
using a three-pinhole interferometer. 
From the interferogram, we can extract the geometric phase related to
 the three-vertex Bargmann invariant as the area of a triangle formed by interference fringes. 
Unlike the conventional methods, our method does not involve the state evolution. 
Moreover, the phase calibration of the interferometer and the elimination of the dynamical
 phase are not required.
The gauge invariance of the geometric phase corresponds to the fact that the
 area of the triangle is never changed by the local phase shift in each internal state. 
\end{abstract}
\maketitle
\textit{Introduction.}---
When a quantum system evolves in time and returns to its 
initial state, the final and initial wavefunctions can differ by a phase
factor composed of two parts: a dynamical phase proportional to 
the time integral of the instantaneous energy and a geometrical phase, 
which depends only on the path traced in the ray
space and not on the energy and the rate of evolution.
The geometric phase was originally discovered by
Berry in the adiabatic and
cyclic evolution of a pure quantum state\,\cite{berry84:_quant_phase_factor_accom_adiab_chang}. 
Since then, it has been generalized to
nonadiabatic evolution\,\cite{aharonov87:_phase_chang_durin_cyclic_quant_evolut,anandan92:_geomet_phase}
and noncyclic evolution\,\cite{samuel88:_gener_settin_for_berry_phase,morinaga07:_berry_phase_for_noncy_rotat}. 
Its applications in practical fields 
such as fault-tolerant quantum computation
\,\cite{zanardi99:_holon_quant_comput,duan01:_geomet_manip_of_trapp_ions}
and weak measurement\,\cite{tamate09:_geomet_aspec_of_weak_measur} are also proposed.

Most approaches for observing the geometric phase, such as 
the interferometric and polarimetric methods
\,\cite{suter88:_study_of_aharon_anand_quant,
kwiat91:_obser_of_noncl_berry_phase_for_photon,
schmitzer93:_nonlin_of_panch_topol_phase,
brendel95:_geomet_phases_in_two_photon_inter_exper,
webb99:_measur_of_berry_s_phase,
loredo09:_measur_of_panch_phase_by,
kobayashi09:_simpl_exper_for_quant_eraser}, 
require the state evolution. 
Note that the geometric phase induced by such evolution appears as 
the global phase factor, which cannot be measured directly.
Therefore in order to observe the geometric phase, 
we must prepare the reference state, i.e., the state that is left
unevolved, and measure
the relative phase between the evolved and the reference states. 
To remove additional phase shifts associated with the above operations, 
phase calibration is
required\,\cite{sjoqvist06:_geomet_phase_in_weak_measur}, 
that is, the relative phase must be
determined by comparing the cases \textit{with} and \textit{without} the
state evolution.
Moreover, the
dynamical phase must be eliminated from the relative phase. 
These two considerations, namely, the phase calibration and the
elimination of the dynamical
phase, lead to experimental complications.

\begin{figure}[b]
\vspace*{-0.4cm}
\includegraphics[width=6.5cm]{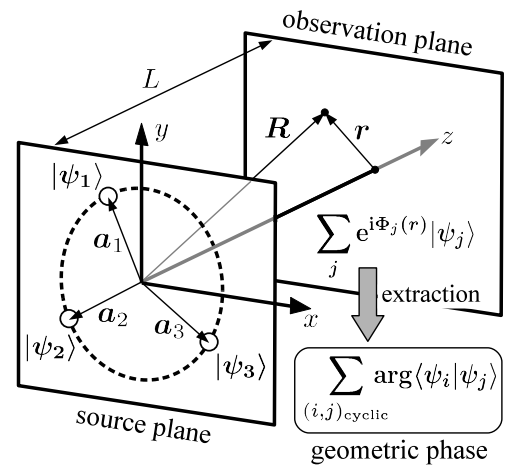}
\vspace*{-0.3cm}
\caption{Coordinate system of three-pinhole interferometer. 
The three pinholes are irradiated by monochromatic light and the
 three spherical waves interfere. 
We can extract the geometric phase for three states,
 $\ket{\psi_1}$, $\ket{\psi_2}$, and $\ket{\psi_3}$, from the
 three-pinhole interferogram.}
\label{fig:3pinhole_interferometer}
\end{figure}

On the other hand, based on the quantum kinematic approach
\,\cite{mukunda93:_quant_kinem_approac_to_geomet_phase}, 
the geometric phase can be attributed to the geometric structure of the
Hilbert space, and not to the state evolution. 
In particular, an important consequence of the kinematic approach has been to show
the close relationship between the geometric phase and the Bargmann
invariant\,\cite{bargmann64:_note_wigner_theor_symmet_operat}. 
The $n$-vertex Bargmann invariant is a complex quantity that is determined
by $n$ points in the ray space.
The phase of the $n$-vertex Bargmann invariant is identical to the geometric
phase for a closed ray-space curve obtained 
by connecting the corresponding $n$ states with geodesics. 
The geometric phase for a
smooth curve can be obtained from the Bargmann invariant by approximating a
smooth curve with a chain of geodesics. 
In this sense, 
the phase of the Bargmann invariant, particularly that of the three-vertex
Bargmann invariant, is the primitive building block of the geometric phase. 

The purpose of this letter is to report a method to measure the
geometric phase or the three-vertex Bargmann invariant 
without the state evolution. 
All we need to do is to prepare the three states and let them interfere
directly. In our method, the phase calibration and elimination of
the dynamical phase are not required.
As shown in Fig.~\ref{fig:3pinhole_interferometer}, 
we assign three states, $\ket{\psi_j}$ ($j=1,2,3$), 
to the internal states of the photon
from three pinholes and obtain a three-pinhole interferogram,
which contains three distinct interference fringes due to each pinhole pair. 
Using a certain data processing,
we can extract the geometric phase directly from the interferogram. 

\textit{Geometric phase and Bargmann invariant.}---
In 1956, Pancharatnam introduced the definition of the phase relation
between any two (non-orthogonal) states 
\,\cite{pancharatnam56:_proc}. 
Assume that $\ket{\psi_1}$ and $\ket{\psi_2}$ are two different
states and $\ket{\psi_1}$ is exposed to the U(1) shift
$\ee^{\ii\phi}$. By superimposing two such states, 
we have the intensity 
$I\propto 1+|\bracketi{\psi_1}{\psi_2}|\cos(\phi+\arg\bracketi{\psi_1}{\psi_2})$. 
The interference fringes are shifted by
$\arg\bracketi{\psi_1}{\psi_2}$, the relative phase between two
states. 
In particular, when they constructively interfere or $\arg\bracketi{\psi_1}{\psi_2}=0$, 
$\ket{\psi_1}$ and $\ket{\psi_2}$ are said to be \textit{in-phase}. 

A remarkable feature
of this relation is its non-transitivity; even if
$\ket{\psi_1}$ is
in-phase with $\ket{\psi_2}$ and $\ket{\psi_2}$ with $\ket{\psi_3}$, the
relative phase between $\ket{\psi_1}$ and $\ket{\psi_3}$ is, in general, not
zero. It is easy to show that the non-zero phase difference
between $\ket{\psi_1}$ and $\ket{\psi_3}$ can be written as
\begin{align}
\Delta_3(\psi_1,\psi_2,\psi_3)&=
\arg\bracketi{\psi_1}{\psi_2}\bracketi{\psi_2}{\psi_3}\bracketi{\psi_3}{\psi_1}\nonumber\\
&=\sum_{(i,j)_\U{cyclic}}\arg\bracketi{\psi_i}{\psi_j},
\label{eq:4}
\end{align}
where $(i,j)_\U{cyclic}\equiv(1,2),(2,3),(3,1)$. 
$\Delta_3$ is called the Pancharatnam
phase\,\cite{pancharatnam56:_proc}. The product 
$\bracketi{\psi_1}{\psi_2}\bracketi{\psi_2}{\psi_3}\bracketi{\psi_3}{\psi_1}$
is the Bargmann
invariant\,\cite{bargmann64:_note_wigner_theor_symmet_operat} 
for the three states. Equation~(\ref{eq:4}) is gauge invariant, i.e.,
independent of the choice of the local phase factor of each state 
because the bra and ket vectors for each state
appear in a pair. The phase
$\Delta_3$ is the primitive building block of the geometric phase
based on the quantum kinematic
approach\,\cite{mukunda93:_quant_kinem_approac_to_geomet_phase}. 
It turns out that $\Delta_3$ is simply related to the geometric phase
associated with the geodesic triangle in the ray space.
For a two-state system such as the
polarization of a photon, $\Delta_3$ is proportional to the solid angle
of the spherical triangle on the Bloch sphere with vertices at
$\ket{\psi_1}$, $\ket{\psi_2}$, and $\ket{\psi_3}$
\,\cite{pancharatnam56:_proc,aravind92:_simpl_proof_of_panch_theor}.

\textit{Geometric phase and ridge lines}---
As shown in Fig.~\ref{fig:3pinhole_interferometer}, 
consider three pinholes irradiated by monochromatic light that has
the wave number $k$. 
Without loss of generality, we may consider that
the three pinholes are located at
$\vct{a}_j$ $(j=1,2,3)$ on the source plane $z=0$ with their origin at the
circumcenter of the triangle formed by the three pinholes, and these vectors
have the same length $a$ 
(see Fig.~\ref{fig:3pinhole_interferometer}). 
The state of each photon from the pinholes is composed of two parts: 
the spatial part that is represented by the spherical wave and 
the internal state of the photon, namely, the polarization
state. Assuming that the transmission probabilities of the three pinholes
are the same for simplicity, the state on an observation plane at a
distance of $L$ is represented by \,\cite{born99:_princ_of_optic}
\begin{align}
\ket{\Psi(\vct{r})}&=
C\sum_{j=1}^3\frac{\ee^{\ii(k|\vct{R}-\vct{a}_j|+\phi_j)}}{|\vct{R}-\vct{a}_j|}\ket{\psi_j},
\label{eq:1}
\end{align}
where $\vct{R}$ is the position vector on the observation plane $z=L$; 
$\vct{r}\equiv\vct{R}-(\vct{R}\cdot\vct{e}_z)\vct{e}_z$, 
the transverse component of $\vct{R}$ with the unit vector $\vct{e}_z$
along the $z$-axis; 
$C$, the dimensionless normalization constant; 
$\phi_j$, the phase  of the $j$th source; and
$\ket{\psi_j}$, the polarization state of the $j$th source. 
$|\vct{R}-\vct{a}_j|$ is the
distance from the $j$th source to a given observation point $\vct{R}$.

We make a paraxial approximation in the far-field regime, and moreover, 
we assume the stronger condition,
$|\vct{r}-\vct{a}_j|\ll(L^3/k)^{1/4}\ll L$.
Under these assumptions, the spherical wave function in Eq.~(\ref{eq:1}) is approximated as
\begin{align}
\frac{\ee^{\ii k|\vct{R}-\vct{a}_j|}}{|\vct{R}-\vct{a}_j|}
\sim \frac{1}{L}\exp\left[\ii k\left(L+\frac{r^2+a^2}{2L}
-\frac{\vct{r}\cdot\vct{a}_j}{L}\right)\right],
\end{align}
where $r\equiv|\vct{r}|$. 
Therefore, the intensity distribution of the interference field, $p(x,y)$, can be
written as 
\begin{align}
p(x,y)&=\|\ket{\Psi(\vct{r})}\|^2\nonumber\\
&=\frac{C^2}{L^2}\Bigl\{-3+\sum_{(i,j)_\U{cyclic}}P_{ij}(x,y)
\Bigr\},
\label{eq:9}
\end{align}
with
\begin{align}
P_{ij}(x,y)&=2\Bigl(1+
|\bracketi{\psi_i}{\psi_j}|\nonumber\\
&\times\cos\bigl[\vct{k}_{ij}\cdot\vct{r}
-\phi_{ij}+\arg\bracketi{\psi_i}{\psi_j}\bigr]\Bigr),
\label{eq:2}
\end{align}
where
$\vct{k}_{ij}\equiv k(\vct{a}_i-\vct{a}_j)/L$ and $\phi_{ij}\equiv\phi_i-\phi_j$.
Equation~(\ref{eq:2}) corresponds to the double-slit interference
fringe between the two states, $\ket{\psi_i}$ and
$\ket{\psi_j}$\,\cite{pancharatnam56:_proc}. 
Therefore, the three-pinhole interferogram (\ref{eq:9}) contains 
three sets of interference fringes with different directions. 

In order to extract the geometric phase from the total interferogram 
(\ref{eq:9}), we should focus our attention on the phase 
of the interference fringes $P_{ij}$, 
since their visibility $|\bracketi{\psi_i}{\psi_j}|$ includes no
information about the geometric phase in Eq.~(\ref{eq:4}).
Here, we consider the phase condition to attain the maximum
of each interference fringe in Eq.~(\ref{eq:2}),
\begin{align}
\vct{k}_{ij}\cdot\vct{r}=\phi_{ij}-\arg\bracketi{\psi_i}{\psi_j}+2n_{ij}\pi,
\label{eq:3}
\end{align}
where $n_{ij}$ are integers. 
Equation~(\ref{eq:3}) defines three distinct sets of parallel lines, which
we call \textit{ridge lines}, on the observation plane $z=L$. 
The area $S$ of the
triangle formed by the three distinct ridge lines, which we call a \textit{ridge
triangle}, is calculated as
\begin{align}
 S=\frac{L^2}{4k^2S_0}\cdot\left\{\Delta_3(\psi_1,\psi_2,\psi_3)-2n\pi\right\}^2,
\label{eq:5}
\end{align}
where $n=n_{12}+n_{23}+n_{31}$ and 
$S_0$ is the area of the triangle formed by the three pinholes. 
Equation~(\ref{eq:5}) shows three important features of a ridge
triangle. First, the area of the ridge triangle is essentially related to
the geometric phase $\Delta_3$.
In particular, we call a ridge triangle that includes no ridge lines
inside as \textit{elemental}. 
Assuming $0\leq\Delta_3<2\pi$, ridge triangles with
$n=0$ and $1$ in Eq.~(\ref{eq:5}) are elemental. The area with $n=0$ is proportional 
to the square of the geometric phase $\Delta_3$
and that with $n=1$ is proportional to the square of $2\pi-\Delta_3$. 
Second, we should note that the area $S$ does not depend on the choice of
the local phases $\phi_i$, because the geometric phase $\Delta_3$ is gauge invariant.
By introducing a phase shift to one of the pinholes, two sets of ridge
lines are displaced but the areas of the ridge triangles are conserved.
Third, any geometry of the three pinholes can form the ridge triangle
related to the geometric phase since Eq.~(\ref{eq:5}) is proportional to
the square of the geometric phase regardless of the
vectors $\vct{a}_i$.

\begin{figure}[tb]
\begin{center}
\includegraphics[width=8cm]{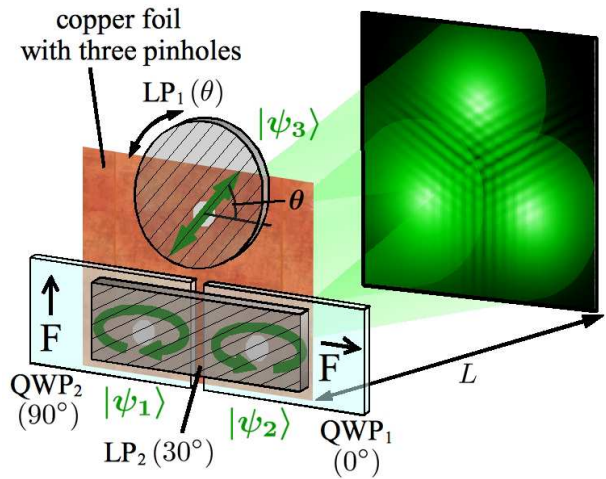}
\end{center}
\vspace*{-0.7cm}
\caption{(color online). 
Experimental setup for three-pinhole interference with
 different polarization states. In front of the upper pinhole, we placed 
a film-type linear polarizer, LP$_1$,
attached to a rotatable mount with graduated scales for adjusting the
angle $\theta$, whereas in front of the lower left and right pinholes we
placed film-type quarter-wave plates having orthogonal fast
axes, $0^\circ$ (QWP$_1$) and $90^\circ$ (QWP$_2$), respectively,
behind a linear polarizer LP$_2$ with a fixed angle of $30^\circ$.
Incident light on the pinholes is circularly polarized, 
and transmittance of light through each pinhole is 50\%. 
Under this configuration, the visibility of the fringe $P_{ij}(x,y)$ 
is more than or equal to 0.5, which is
sufficient to retrieve clear ridge lines.
}
\label{fig:three_beam}
\end{figure}
\begin{figure}[tb]
\begin{center}
\includegraphics[width=9cm]{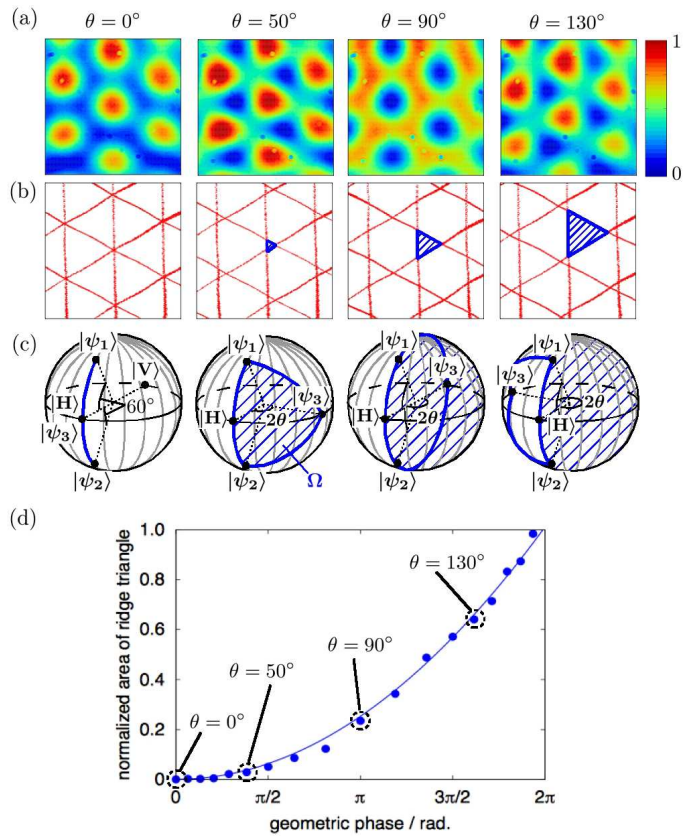}
\end{center}
\vspace*{-0.5cm}
\caption{(color online). 
Ridge triangles and geometric phases 
for $\theta=0^\circ$, $50^\circ$, $90^\circ$, and $130^\circ$. 
(a) The three-pinhole interferogram obtained in our experiment. 
(b) Ridge lines extracted from the above interferograms. 
(c) Corresponding spherical triangle on the Poincar\'{e} sphere. 
(d) Geometric phase versus area of ridge triangle. The area of the ridge
 triangle is normalized by the maximum area. The solid line indicates the
 theoretical curve, and it shows a quadratic characteristic.}
\label{fig:interference_pattern_and_ridge}
\end{figure}

\textit{Extraction of ridge lines}---
A straightforward method to determine the ridge lines is 
the observation of individual interference fringes
$P_{ij}(x,y)$ in Eq.~(\ref{eq:2}) by closing one of the three pinholes. 
However, instead of using the three interferograms $P_{ij}(x,y)$, we can
extract all the ridge lines from a single-shot interferogram $p(x,y)$ with the
three pinholes.
First, we introduce the vectors
\begin{align}
\vct{b}_i&\equiv
\vct{e}_z\times\left(\vct{a}_j-\vct{a}_k\right)
=\frac{L}{k}\vct{e}_z\times\vct{k}_{jk},
\end{align}
where $(i,j,k)=(1,2,3),(2,3,1),(3,1,2)$. 
The vector $\vct{b}_i$ on the observation plane $z=0$ is determined only
from the geometry of the three pinholes.
Considering the directional derivative along $\vct{b}_i$, we can
eliminate one of the interference fringes, $P_{jk}(x,y)$, from the total
interferogram $p(x,y)$ since $\vct{b}_i$ is orthogonal to
$\vct{k}_{jk}$. The other two fringes remain as sinusoidal functions. 
In addition, by applying another directional derivative along $\vct{b}_j$, 
we can isolate the oscillation term of $P_{ij}(x,y)$ from
$p(x,y)$ as
\begin{align}
\left(\vct{b}_i\cdot\vct{\nabla}\right)&\left(\vct{b}_j\cdot\vct{\nabla}\right)p(x,y)
\nonumber\\
\propto&|\bracketi{\psi_i}{\psi_j}|\cos\bigl[\vct{k}_{ij}\cdot\vct{r}
-\phi_{ij}+\arg\bracketi{\psi_i}{\psi_j}\bigr].
\label{eq:10}
\end{align}
Then, a set of the ridge lines can be retrieved from Eq.~(\ref{eq:10}).
Examples are shown in
Fig.~\ref{fig:interference_pattern_and_ridge}. Interferograms for
the three pinholes are shown in Fig.~\ref{fig:interference_pattern_and_ridge}(a) and 
the three sets of ridge lines thus extracted 
are shown in Fig.~\ref{fig:interference_pattern_and_ridge}(b).

As a result, we can determine the pure geometric phase instantaneously as the square root of
the area of the ridge triangle extracted directly from the three-pinhole
interferogram for three arbitrary states.

\textit{Experiments.}---
Our experimental setup is shown in
Fig.~\ref{fig:three_beam}. 
The light source is a 532-nm green laser. 
The source illuminates a thin copper foil that is perforated with three
0.1-mm-radius pinholes forming an equilateral triangle of side
length $1.5\,\U{mm}$.
At a distance of approximately $2\,\U{m}$ from the three pinholes, the
interfering patterns are captured using a charge-coupled device (CCD) camera. 
The CCD camera has an image resolution of $640\times 480$ pixels, 
with the size of each pixel being $9\,\U{um}\times 8\,\U{um}$.
According to the setting of polarization elements in Fig.~\ref{fig:three_beam}, 
the polarization states from the left, right, and upper pinholes are
$\ket{\psi_1}=(\sqrt{3}\ket{\U{H}}+\ii\ket{\U{V}})/2$,
$\ket{\psi_2}=(\ii\sqrt{3}\ket{\U{H}}+\ket{\U{V}})/2$, and
$\ket{\psi_3}=\cos\theta\ket{\U{H}}+\sin\theta\ket{\U{V}}$,
where $\ket{\U{H}}$ and $\ket{\U{V}}$ are the horizontal and 
vertical polarization states, respectively.

On a Poincar\'{e} sphere, 
$\ket{\psi_1}$ and $\ket{\psi_2}$ are both 
located at a latitude of $\pm 60^\circ$ on the prime meridian, 
and $\ket{\psi_3}$ is located on
the equator at a longitude of $2\theta$, which can be varied according
to the setting of LP$_1$. 
The geometric phase is proportional to the solid angle $\Omega$ of the spherical
triangle formed by $\ket{\psi_1}$, $\ket{\psi_2}$,
and $\ket{\psi_3}$ on the Poincar\'{e} sphere, i.e.,
$\Delta_3=-\Omega/2$. It is calculated as
\begin{align}
\Delta_3(\psi_1,\psi_2,\psi_3)=\tan^{-1}\left(\frac{1}{\sqrt{3}}\tan\theta\right),
\label{eq:6}
\end{align}
which moves between $0$ and $2\pi$ with respect to $\theta$.

Figure~\ref{fig:interference_pattern_and_ridge} shows our experimental
results. 
Figure~\ref{fig:interference_pattern_and_ridge}(a) shows 
the experimentally obtained interferograms for several values of $\theta$, 
and Fig.~\ref{fig:interference_pattern_and_ridge}(b), the ridge lines
extracted from the above interferograms. The shaded triangles
in Fig.~\ref{fig:interference_pattern_and_ridge}(b) are the elemental
ridge triangles ($n=0$) and the area of the ridge triangle 
varies with the spherical triangle on the Poincar\'{e} sphere 
[see Fig.~\ref{fig:interference_pattern_and_ridge}(c)]. 
The relationship between the elemental ridge triangle and the geometric
phase is quantitatively analyzed in
Fig.~\ref{fig:interference_pattern_and_ridge}(d), in which the area of the
elemental ridge triangle normalized by the maximum area is plotted as a function of the
geometric phase calculated from Eq.~(\ref{eq:6}). 
The solid line in Fig.~\ref{fig:interference_pattern_and_ridge}(d) is
the theoretical curve calculated from Eq.~(\ref{eq:5}), and the experimental results 
[dots in Fig.~\ref{fig:interference_pattern_and_ridge}(d)] are found to agree well with
the theoretical prediction. 
In Fig.~\ref{fig:interference_pattern_and_ridge}(b), 
we can also see the other elemental ridge triangle ($n=1$), which is
related to the complementary area on the Poincar\'{e} sphere
$4\pi-\Omega$.

\begin{figure}[tb]
 \begin{center}
\includegraphics[width=8.5cm]{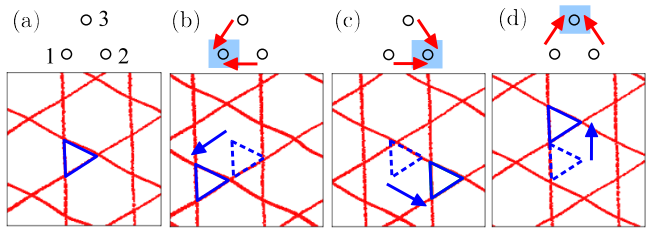}
 \end{center}
\vspace*{-0.5cm}
\caption{(color online). 
Ridge triangles for $\theta=90^\circ$ with (a) no
 shifts, (b) shift on pinhole 1, (c) shift on pinhole 2, and
(d) shift on pinhole 3. The ridge triangle
 is parallel displaced, but it is not deformed for a local phase shift.}
\label{fig:delayed_ridge}
\end{figure}
Figure~\ref{fig:delayed_ridge} shows variations of the ridge triangles when 
a local phase shift is introduced by inserting 
a thin (0.15-mm-thick) glass plate in front
of each pinhole. Figure~\ref{fig:delayed_ridge}(a) shows the ridge lines
without a phase shift as reference.
When a phase shift is introduced at pinhole 1, as shown in
Fig.~\ref{fig:delayed_ridge}(b), 
the two interference fringes $P_{12}(x,y)$ and $P_{31}(x,y)$ in
Eq.~(\ref{eq:2}) suffer the same phase shift and they are simultaneously displaced
toward pinhole 1. Thus, the ridge triangle is only
parallel displaced along the ridge line of fringe $P_{23}(x,y)$, but it is
not deformed. Similarly,
a phase shift applied to pinhole 2 and pinhole 3 has no influence on the
size of the ridge triangle, as shown in Figs.~\ref{fig:delayed_ridge}(c)
and (d), respectively.
The fact that the ridge triangle is not deformed shows 
the gauge invariance of the geometric phase in our experiment.

\textit{Conclusion.}---
We have shown a procedure for measuring
the geometric phase without state evolution 
using a three-pinhole interferometer. 
From the interferogram, we can extract the primitive building block of the
geometric phase $\Delta_3$ 
(phase of the three-vertex Bargmann invariant) 
as the area of the ridge triangle. 
Our experiment requires no procedures for phase calibration and
elimination of the dynamical phase. 
The gauge invariance of the geometric phase corresponds to the fact that
the ridge triangle is not deformed by the local phase shift. 
Moreover, using a CCD video camera followed by image processing to
extract the ridge lines, we can \textit{see} the geometric phase in real time.

We thank Yosuke Nakata for providing useful comments and suggestions.
This research is supported by the global 
COE program ``Photonics and Electronics Science and Engineering'' at Kyoto University.

\end{document}